\magnification=\magstep1
\font\fta=cmr10 scaled\magstep2
\centerline{\fta Exact treatment of Pauli exclusion effects}
\bigskip
\centerline{\fta in (p,p') reactions}
\vskip.35in
\centerline{E.J. Stephenson}
\centerline{\it Indiana University Cyclotron Facility, Bloomington, Indiana 47408}
\bigskip
\centerline{R.C. Johnson}
\centerline{\it Department of Physics, University of Surrey, Guildford, Surrey GU2 5XH, UK}
\bigskip
\centerline{F. Sammarruca}
\centerline{\it Department of Physics, University of Idaho, Moscow, Idaho 83844}
\vskip.4in
\noindent
{\bf Abstract:}~~This paper presents the first calculations of proton inelastic scattering
in which the medium effect of Pauli blocking is included through an exact, rather than angle
averaged, operator.  This improvement is important in the isoscalar channel at proton
energies near 100~MeV and fades as the energy rises.  However, processes that emphasize
finite-range exchange (such as $0^+\to 0^-$ reactions) still see
significant effects at 200~MeV.  The results depend on the directions
of the incident and struck nucleon momenta that produce the most
important contributions to the DWBA integral at each scattering
angle.
\vskip.3in
\noindent
{\bf I.~~Introduction}
\medskip
Proton elastic and inelastic scattering at energies above 100~MeV are usually described by
distorted-wave calculations based on an effective nucleon-nucleon (NN) interaction.  When
the interaction with the projectile proton is summed, or ``folded,'' over all the nucleons
in the target, the resulting potential can be used to describe elastic scattering.  In
addition, the same effective NN interaction becomes the transition potential that connects to
excited states of the target when the struck nucleon moves to a new shell-model orbit,
creating a particle-hole pair.  The
many-body effects of the nuclear medium are
usually incorporated through modifications to this
effective NN interaction that depend on the local nuclear density.

Systematic studies [1--5] have
shown that one important part of the many-body effects is
Pauli blocking, particularly at the lower energies in this range.
In this case, a projectile nucleon travelling through the nuclear medium experiences a
potential that arises from virtual NN scattering, but only to intermediate states for which both
of the nucleons involved have momenta above the Fermi momentum [6].  This
restriction is included through a projection operator in
the Bethe-Goldstone equation for the $G$-matrix elements that describe
the effective NN interaction inside the nuclear medium.

The usual practice is to average this Pauli projection operator over the intermediate state
scattering angle.  It has been argued that this spherical approximation is adequate for
the central and spin-orbit terms in the effective interaction [7].  If this approximation
is removed, new $G$-matrix elements appear [7--10].  While remaining diagonal in total
spin $S$ and isospin $T$, the angular dependence allows coupled $G$-matrix elements
that connect partial wave states where
$J\neq J'$ and $\ell\neq\ell '$ beyond the $\vert\ell -\ell '\vert =2$ coupling
required by the tensor interaction.

In a recent calculation of
the $G$-matrix [10], we observed that these new couplings in $J$ and $\ell$ generate
only small matrix elements, but the variation of $M$,
the projection of the total angular momentum $J$,
produces changes in the $G$-matrix elements that are similar in size
to the Pauli blocking effect inself.  Thus it seems appropriate to inquire
whether it is possible to observe these differences between the exact and
spherical Pauli operators in selected nuclear reactions, and in particular in
their spin observables.  Proton inelastic scattering offers a rich set of
polarization observables, especially considering polarization transfer, that reflect the spin
dependence of the effective NN interaction itself.
Calculations that have considered only how the binding energy of
nuclear matter is affected by the change from a spherical to an exact treatment
of Pauli blocking show very small changes [8,9].

The dependence on $M$ has been considered previously in
calculations of deuteron binding in the presence of the nuclear medium [11].
While Pauli blocking reduces the deuteron binding energy, the amount depends on whether
the projection of the deuteron's spin 1, $\vert M\vert$, is 0 or 1 when the quantization axis
is taken along the direction of motion of the deuteron through nuclear matter.  This
difference generates
a momentum-dependent $T_P$ tensor potential in the deuteron-nucleus optical
model.  Optical model calculations
have shown that $T_P$ effects are best distinguished from
those of a $T_R$ tensor potential when the deuteron elastic scattering angular distribution
is far-side dominated.
In this case the orbiting trajectories make the deuteron's momentum
nearly perpendicular to the radius from the center of the target when the deuteron is
in the region of the nuclear surface [12].
Measurements of observables such as $X_2=
(2A_{xx}+A_{yy})/\sqrt{3}$ where spin-orbit
effects are suppressed [13] are still insufficient to demonstrate
unambiguously the existence of such a $T_P$
potential, in part because it is necessary to also consider breakup channel coupling and the
necessary computer programs to incorporate both do not exist.

In this paper, we begin with the $G$-matrix described in Ref.~[10].  The important features
of the calculation of the $G$-matrix elements
are reviewed in Section~II where we point out the aspects crucial for
the reaction calculations.  In order to use this effective interaction in distorted-wave
calculations with presently-available computer programs,
we must transform the matrix elements to a coordinate-space representation
using a sum of Yukawa functions.  In contrast to the single-step transformation
described in Ref.~[5], we will first convert the partial-wave matrix elements to angular
distributions of the NN scattering amplitudes (see Section III).  In the process, we will
introduce an expansion of the new $G$-matrix elements in which the lowest order recovers the
result for a spherical Pauli blocking operator.  The coefficients of the Yukawa
expansion will then be determined from a fit to these amplitudes.
For the spherical Pauli operator, this produces the same result as the
previous method [5].
In general, the new
elements of the $G$-matrix require a larger range of spin operators than normally appears
for NN scattering.  Since these new operators are not available in existing distorted-wave
programs, we did not include their contributions to proton-induced inelastic scattering.

In these calculations, it is natural to quantize along the direction of the
momentum of the system.  In nuclear matter, this is the momentum of the
projectile, as noted earlier for the case of the deuteron.  For this choice,
large differences were seen previously for $g$-matrix elements with
different values of $M$ [10].  For the effective interaction that enters
into elastic and inelastic scattering, it is also possible to choose as
the system momentum the sum of the momenta of the incident and struck
nucleons.  For a reaction whose product is observed at a particular
scattering angle, it is no longer required by symmetry that this sum
average to the projectile direction.  The question of what is the most
important direction for this sum also arises when combining the direct
and exchange parts of the effective NN interaction when making a
zero-range DWIA calculation.
In Appendix B of Ref.~[5] we presented a scheme
in which the struck nucleon's momentum
is chosen so that, at any given momentum transfer (associated with a
particular scattering angle), the scattering has on-shell kinematics.
In the limit where the reaction Q-value vanishes and the recoil of the
target is neglected, this scheme places the momentum of the two-nucleon
system (sum of the incident and struck nucleon momenta) pointing in the
reaction plane at an angle that is half of the scattering angle.  Thus
the best choice for nucleon-induced reactions may not be the same as
for nucleons travelling through nuclear matter.  As part of the
development described here, we will explore both
of these options as a test of
the importance of this issue of kinematics for the treatment of Pauli
blocking.

In Section III we will compare the exact and spherical treatments of the Pauli blocking
operator for representative nuclear transitions at 100 and 200~MeV.  We will include a
comparison to the free, or density-independent, effective interaction to help gauge the
importance of the exact treatment relative to the effect of not including Pauli blocking
at all.  We will show that this relative importance rises at lower bombarding energies.

We will also compare with measurements of the cross section and analyzing power to
illustrate these new effects in relation to the quality of the reproduction of these data.
Pauli blocking has its largest effects on the isoscalar central and spin-orbit terms in
the effective interaction.  These terms are well tested by a comparison to elastic proton
scattering or transitions to natural-parity excited states.
Since the two alternatives for the choice of the system momentum arise out of a
consideration of exchange between the projectile and
the struck nucleon, we will include calculations for a $0^+\to 0^-$ transition.  In
this spin structure there is no analyzing power (in a plane-wave calculation) unless there
is finite-range exchange in the distorted-wave calculation.  So here we
might expect to be sensitive to this issue.

Pauli blocking is only one process that is important in the calculation of the effective
interaction in the nuclear medium.  Others, such as the effects of strong relativistic
mean fields [10,14,15] and coupling to $\Delta$-resonances [16], increase the repulsion in
the nuclear medium just as does Pauli blocking [17].
In a complete treatment, these should be properly
considered,
along with the attraction expected to arise from many-body forces.
 Thus we will not expect that a set of calculations that includes only
Pauli effects will produce good agreement with the data.
However, any critical evaluation of any of
these medium effects requires that the treatment of Pauli blocking not introduce systematic
errors large enough to affect our interpretation when agreement with data is considered.
We will show that in this energy range one must include an exact treatment of the angular
dependence of the blocking operator in order to meet this standard.
\vskip.3in
\noindent{\bf II.~~Calculation of the $G$-matrix elements}
\medskip
The Brueckner-Bethe-Goldstone equation [18--21] describes scattering of two nucleons in nuclear
matter. The presence of the (infinite) medium is accounted for through Pauli blocking
and a mean field arising from the interactions with all the other nucleons.
It is convenient to express the momenta of the two nucleons, ${\bf k}_1$ and
${\bf k}_2$, in terms of the relative and center-of-mass motion as
$${\bf k}=({\bf k}_1-{\bf k}_2)/2\eqno{(1a)}$$ $${\bf P}=({\bf k}_1+{\bf k}_2)/2\ .
\eqno{(1b)}$$  The total or center-of-mass momentum {\bf P} is conserved in the
scattering process.

In strict analogy with free-space scattering, the
Bethe-Goldstone equation is given by
$$ G(\vec q',\vec q,\vec P,E_0) = V(\vec q',\vec q) \eqno{(2)}$$
$$ + \int {d^3K\over(2\pi)^3}V(\vec q',\vec K)
{Q(\vec K,\vec P)\over E_0 -E(\vec P,\vec K)}
G(\vec K,\vec q,\vec P,E_0)\ ,$$
where $V$ is the two-body potential.
 The energy of the two-particle system, $E$ (with $E_0$ its initial value), includes
kinetic energy and the potential energy generated by the mean field. The latter
is determined in a separate self-consistent
calculation of nuclear matter properties and conveniently parametrized in terms of effective
masses [5].

The Pauli projection operator $Q$ selects intermediate states only when both momenta
lie above the Fermi momentum $k_F$:$$Q({\bf k},{\bf P},k_F)=\cases{1&if $k_1,k_2>k_F$\cr
0&otherwise\cr}\eqno{(3)}$$
Visualizing the (sharp) Fermi surface as a sphere of radius $k_F$, the condition above imposes the requirement
that the tips of the ${\bf k}_1$ and ${\bf k}_2$ vectors lie outside the sphere.
For applications to real nuclei, $k_F$ is treated as a function of the
local nuclear density $$\rho(r)={2k_F^3(r)\over 3\pi^2}\ .\eqno{(4)}$$

The matrix elements for the exact Pauli operator may be written in a partial wave basis
as $$\langle (\ell'S)J'M\vert Q(k,P,k_F)\vert (\ell S)JM\rangle\eqno{(5)}$$
$$=\sum_{m_\ell ,m_S}
\langle\ell 'm_\ell Sm_S\vert J'M\rangle\langle JM\vert\ell m_\ell Sm_S\rangle
\langle\ell'm_\ell\vert Q(k,P,k_F)\vert\ell m_\ell\rangle\ ,$$ where $$\langle
\ell'm_\ell\vert Q(k,P,k_F)\vert\ell m_\ell\rangle =\int d\Omega\ Y^*_{\ell'm_\ell}
(\Omega )\ Y^{}_{\ell m_\ell}(\Omega )\ \Theta (\vert{\bf k}_1\vert -k_F)\Theta (\vert
{\bf k}_2\vert -k_F)\ .\eqno{(6)}$$  The step functions destroy the orthogonality
that would otherwise exist for the spherical harmonics in the integral.  This allows
couplings to appear where $\ell\neq\ell '$ and, through the recoupling coefficients
in the summation, to couplings where $J\neq J'$.
Since the step functions involve the polar angle $\theta$ and not the azimuthal angle
$\phi$, mixing does not arise between different values of $m_\ell$ and $m_{\ell'}$, and
this carries over into the values of $M$ and $M'$.  This restriction is already
incorporated into Eqs.~(5) and (6).  Thus we finally must consider a
$G$-matrix element whose general spin and isospin structure is $$G=\langle\ell'J'\vert
G^{ST}_M\vert\ell J\rangle\ .\eqno{(7)}$$

The $G$-matrix elements that we will use here
were generated from a free-space NN interaction that is a
modified version of the Bonn-B potential [17].  Pseudo-vector coupling is used for the
pion, and the $\sigma$-meson coupling is allowed to assume different values as a function
of isospin and vary over a limited range in the lowest partial waves.  The masses, coupling
constants, and cutoff parameters may be found in Ref.~[5].  All of these parameters
of the force were adjusted to match
the phase shift analysis results from the Nijmegen group at all energies
up to 325~MeV [22].

The number of coupled channels in Eq.~(7) increases with $M$ as it becomes possible to incorporate
larger values of $J$.  However, the $M$-dependence decreases with
increasing $J$ [10] and we were
able to ignore medium effects on partial waves where $J,J'>6$ because of the small size
of these effects.
\vskip.3in
\noindent
{\bf III.~~Transformation of the $G$-matrix to coordinate space}
\medskip
We will begin this section with a description of the general formulas that connect the
expanded $G$-matrix of Eq.~(7) to NN scattering amplitudes, regardless of the complexity
of the coupling.  Next, we will summarize the
simplified forms used for the
calculations reported here, explaining in each case what features have been left out.

The $G$-matrix elements of Eq.~(7), $\langle\ell 'J'M\vert G^{ST}(\vec P)\vert\ell JM
\rangle$,
may be expanded as a function of $L$ where $L$ is the angular
momentum that recouples $J$ to $J'$ using the coefficients $G^{LST}(\ell 'J',\ell J,P)$.  While the $z$-axis in Eq.~(6) was taken to lie along the
projectile momentum and thus $\vec P$ did not appear in Eq.~(7), here
we include $\vec P$ explicitly to generalize this result for reactions.
Thus, $$\langle\ell'J'M\vert G^{ST}(\vec P)
\vert\ell JM\rangle =\sqrt{4\pi}\sum_{L\Lambda}\langle J'M,L\Lambda\vert JM\rangle\ \hat L
\ Y_{L\Lambda}(\vec P)\ G^{LST}(\ell 'J',\ell J,P)\ .\eqno{(8)}$$
Equation (8) can be inverted
to yield the expansion coefficients $$G^{LST}(\ell 'J',\ell J,P)
={1\over\hat J^2}\sum_M\langle
J'M,L0\vert JM\rangle\ \langle\ell 'J'M\vert G^{ST}(\vec P)\vert\ell JM\rangle\ .\eqno{(9)}$$

Since $\vec P$ is the average momentum of the colliding nucleons, it is invariant under
their interchange.  So the anti-symmetrized $G$-matrix element is obtained by subtracting
from Eq.~(9) the same terms that are there but with an additional phase of $(-)^{\ell +
S+T}=-1$, leaving the set that normally describes NN scattering.  $L$ takes on only even values.

We would now like to convert the representation from partial wave angular momenta to
amplitudes as a function of angle.  In the process, we wish to separate the parts of $G$
associated with the central, spin-orbit, and tensor operators usually used to describe
the spin structure of the
NN scattering amplitudes.  In particular, we want to consider the form
$$G^{LST}_{}=G^{LST}_C+
\delta_{S1}\bigl[ G^{L1T}_{LS} (\vec\sigma_1+\vec\sigma_2)\kern-2pt\cdot\kern-2pt\hat n+
G^{L1T}_{TD}S^{}_{12}(\hat q)+G^{L1T}_{TX}S^{}_{12}(\hat Q)\bigr]\eqno{(10)}$$ where $\vec
{\bf q}=\vec{\bf k}'_1-\vec{\bf k}_1$ is the momentum transfer, $\hat n$ is the normal
to the scattering plane, and $\hat Q=\hat q\times\hat n$.
Each of the $G^{LST}_i$ is a function of angle.
The subscript indicates the
part of the NN amplitude, using $C$ for central, both $S=0$ and $S=1$, $LS$ for spin-orbit,
and $TD$ and $TX$ for the ``direct'' and ``exchange'' parts of the tensor interaction.

The coefficients of Eq.~(10) are simply related to
an expansion in spherical harmonics where the coefficients
$G^{ST}_{kq}(\vec k,\vec k',L)$ can be obtained from the
coefficients of of Eq.~(9) for all spin operators through
$$G^{ST}_{kq}(\vec{\bf k},\vec{\bf k}',L)={\rm Trace}\ (G^{LST}_{}
\tau^{}_{kq})\eqno{(11)}$$ where $$\langle S\sigma '\vert\tau_{kq}\vert S\sigma\rangle =
\hat k\ \langle S\sigma ,kq\vert S\sigma '\rangle\eqno{(12)}$$ and $k$ runs from 0 to
$2S$.  For the coefficients shown in
Eq.~(10), $$\eqalignno{G^{LST}_C=&\hat S^{-2}\ G^{ST}_{00}
(\vec k,\vec k',L)&(13)\cr G^{L1T}_{LS}=&{\sqrt{2\pi}\over 6}\sum_qY^*_{1q}(\vec n)\
G^{1T}_{1q}(\vec k,\vec k',L)&(14)\cr G^{L1T}_{TD}-{1\over 2}G^{L1T}_{TX}=&{\sqrt{10\pi}
\over 30}\sum_qY^*_{2q}(\vec q)\ G^{1T}_{2q}(\vec k,\vec k',L)&(15)\cr -{1\over 2}G^{L1T}
_{TD}+G^{L1T}_{TX}=&{\sqrt{10\pi}\over 30}\sum_qY^*_{2q}(\vec Q)\ G^{1T}_{2q}(\vec k,
\vec k',L)\ .&(16)\cr}$$

The $G^{ST}_{kq}(\vec k,\vec k',L)$ are functions of the scattering angle through the
directions of $\vec k$ and $\vec k'$.  They are
obtained from $$\eqalign{G^{ST}_{kq}(\vec k,
\vec k',L)=&\sum_\zeta\hat S\hat J^2\hat J'\hat k_\ell\hat L(-1)^{J-J'-\ell +k_\ell}
\left\{ \matrix{\ell &\ell '&k_\ell\cr J&J'&L\cr S&S&k\cr}\right\} \cr \times &
\langle k_\ell q_\ell ,L\Lambda\vert kq\rangle\ \langle\ell 'm_{\ell '},\ell m_\ell
\vert k_\ell q_\ell\rangle\ Y_{\ell 'm_{\ell '}}(\vec k')\ Y_{\ell m_\ell}(\vec k)\cr
\times &\sqrt{4\pi}\ Y_{L\Lambda}(\vec P)\ G^{LST}_{}(\ell 'J',\ell J,P)\cr}
\eqno{(17)}$$
where the sum runs over $\zeta =\ell 'J'\ell Jm_{\ell '}m_\ell k_\ell q_\ell\Lambda$.

In our present application, these general formuli can be simplified by making specific
choices of the coordinate system for the description of the scattering angles.
For a right handed coordinate system with $\hat z$ along the projectile direction $\hat k$,
$\hat y$ along $\hat n$, and the momentum $\vec P$ in the scattering plane
at an angle $\theta_P$,
Eq.~(17) reduces to $$\eqalign{G^{ST}_{kq}(\vec k,\vec k',L)
=&\sum_\zeta\hat S\hat J^2\hat J'\hat k_\ell\hat L(-1)^{J-J'-\ell +k_\ell}\left\{
\matrix{\ell &\ell '&k_\ell\cr J&J'&L\cr S&S&k_S\cr}\right\} \cr \times &\langle k_\ell
q_\ell ,L\Lambda\vert kq\rangle\ \langle\ell 'q_\ell ,\ell 0\vert kq\rangle\ Y_{\ell '
q_\ell}(\theta ,0)\ \hat\ell\ Y_{L\Lambda}(\theta_P,0)\cr \times &G^{LST}(\ell 'J',
\ell J,P)\cr}\eqno{(18)}$$

The values of $G^{LST}$ were calculated using $$\eqalign{G^{LST}(\ell J\theta_P)=&{1\over
\hat J^2}\ \Bigl\{\langle J0,L0\vert J0\rangle\ \langle\ell JM=0\vert G^{ST}(P)\vert
\ell JM=0\rangle\cr +&\sum_{M>0}\bigl(\langle JM,L0\vert JM\rangle +\langle JM,L0\vert
J-M\rangle\bigr)\ \langle\ell JM\vert G^{ST}(P)\vert\ell JM\rangle\Bigr\}\cr}
\eqno{(19)}$$  The angle-dependent transform was calculated as $$\eqalign{G^{LST}_{kq}=&
\hat L\sum_{J\ell}\hat J^3(-1)^\ell\ G^{LST}_{}(\ell J\theta_P)\sum_{k_\ell q_\ell
\Lambda}\hat k_\ell (-1)^{k_\ell}\ \left\{\matrix{\ell &\ell '&k_\ell\cr J&J&L\cr S&S&k_S
\cr}\right\}\cr \times &\langle k_\ell q_\ell ,L\Lambda\vert kq\rangle\ \langle\ell '
q_\ell ,\ell 0\vert k_\ell q_\ell\rangle\ Y_{\ell 'q_\ell}(\theta ,0)\ Y_{L\Lambda}
(\theta_P,0)\cr}\eqno{(20)}$$ and $$\eqalignno{G^{LST}_C=&\hat S^{-2}\ G^{ST}_{00}(
\vec k,\vec k',L)&(20)\cr G^{L1T}_{LS} =&{i\sqrt{3}\over 6}\ G^{1T}_{11}(\vec k,\vec k',
L)&(21)\cr G^{L1T}_{TD}=&{\sqrt{2}\over 72}\ \Bigl[ 3(1-\cos\theta )G^{1T}_{20}+2\sqrt{6}
\sin\theta\ G^{1T}_{21}+\sqrt{6}(3+\cos\theta )G^{1T}_{22}\Bigr] &(22)\cr G^{L1T}_{TX}
=&{\sqrt{2}\over 72}\ \Bigl[ 3(1+\cos\theta )G^{1T}_{20}-2\sqrt{6}\sin\theta\ G^{1T}_{21}
+\sqrt{6}(3-\cos\theta )G^{1T}_{22}\Bigr]\ .&(23)\cr}$$  An additional transform was needed
at the end to replace states where $S=0,1$ and $T=0,1$ with the singlet and triplet spin
and isospin operators customarily used by the distorted-wave programs.

In Ref.~[10], it was noted that the spherical Pauli operator gave results very
close to the average over $M$ values.  The amplitudes obtained with $L=0$ only also
reproduce this same result.

Values of the density-dependent $G$-matrix calculated with the
spherical Pauli operator were used
for entries where
$7\leq J\leq 15$, matrix elements that we needed to specify the long-range pion tail of
the NN interaction.  No values of $J>15$ were considered.

The amplitudes were then calculated at a number of values of $\theta$ and reproduced using
a sum of Yukawa functions [5,23].  The matrix inversion scheme outlined in Ref.~[5] again
produces the solution in a single step.

Calculations were made for both $\theta_P=0$ (along $\hat z$) and $\theta_P=\theta /2$.
When the $L=2$ terms were included
in the NN scattering amplitudes for $\theta_P=\theta /2$, it was no longer possible to
produce a high quality fit with the usual number of Yukawa coefficients.  Instead, the
direct and exchange parts of the interaction were allowed to have separate coefficients.
In practice, the coefficient values for the direct and exchange expansions were close.
Because the $L=2$ contributions were much smaller than for $L=0$, only $L=2$ was
considered and terms with $L\geq 4$ were ignored.
\vskip.3in
\noindent
{\bf IV.~~Results for (p,p$'$) reactions}
\medskip
\midinsert
\vbox to 3.6in{\vss \hbox to 6.5truein{
\includegraphics{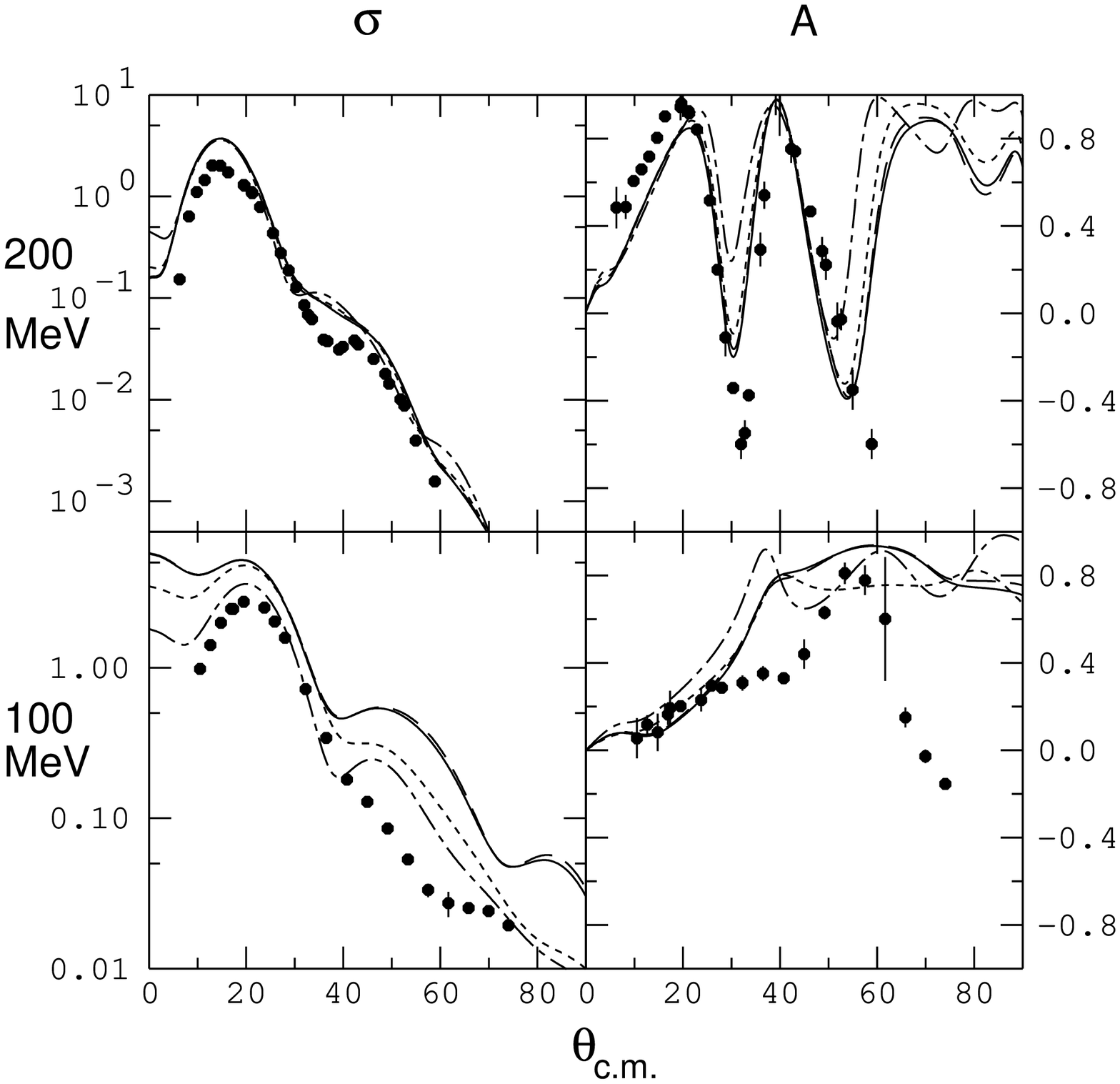}
\hss}}
\noindent
{\it Figure 1:}~~Measurements of cross section and analyzing power for the transition to
the first $2^+$ state at 6.917~MeV in $^{16}$O at two bombarding energies, 200 and 100~MeV.
The data are from Refs.~[24,25].  The calculations use free (dash-dot), spherical Pauli
(short dash), and exact Pauli interactions with $\vec P$ along $\hat z$ (long dash)
or $\theta /2$ (solid).
\endinsert

In this section we will compare (p,p$'$) calculations made with the two forms of the
exact Pauli projection operator (quantized along $\hat z$ or $\theta /2$) and
representative sets of data.  Figure~1 contains the cross section and analyzing power
for the first $2^+$ state in $^{16}$O at 6.917~MeV.  The 200-MeV data are from
Ref.~[24]; the 100-MeV data are reported in Ref.~[25].  The calculations were made with
the distorted-wave program LEA [26].  The transition formfactors reproduce the
inelastic electron scattering measurements of Buti [27].  The distorted waves were
calculated from a
folded optical potential based on the same density-dependent interaction that is used
for the transition.  (Relativistic effects as described in Ref.~[5] are not included.)

The changes made by including medium effects represented by the spherical Pauli blocking
operator as compared with no medium effect is represented
by the difference between the short-dashed (spherical Pauli) and dash-dot (free) curves.
At 200~MeV, changes are noticeable only at the largest angles in the analyzing power.
Including Pauli blocking increases the size of the
diffractive oscillations, a change that goes
in the direction toward better agreement.  If the angle averaged operator is replaced by
the exact operator, only small additional changes result [medium dash (solid)
has $\vec P$
along $\hat z$ ($\theta /2$)].  There is almost no effect from the choice of
system momentum.

At 100~MeV, the Pauli blocking medium effects are larger.  In this case
the change from the spherically averaged to the exact operator makes a much larger
difference, especially for the cross section.  This is in keeping with the expectation
that as the projectile momentum goes down, a greater fraction of the scattered states
will be eliminated from consideration by the step functions of Eq.~(5), and the
procedure for doing this will matter more.  Compared to the change wrought by using
the exact operator, the choice of system momentum still appears not to matter.
This test
demonstrates that, especially near 100~MeV, the exact treatment of the Pauli blocking
operator is important for an accurate description of medium effects.

\midinsert
\vbox to 3in{\vss\hbox to 6.5truein{
\includegraphics{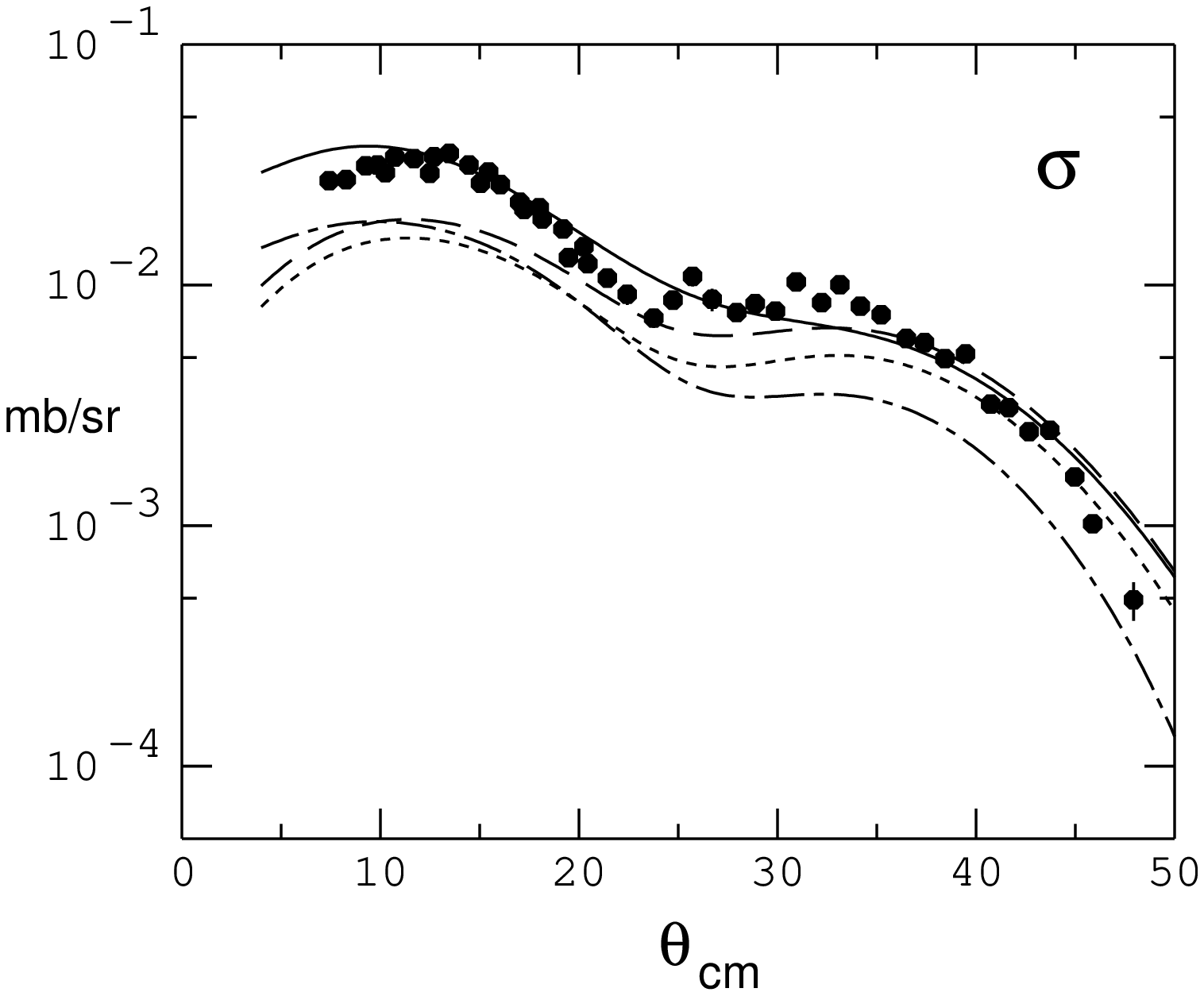}
\hss}}
\noindent
{\it Figure 2:}~~Measurements of the cross section for the transition to the $T=0$, $0^+$
state at 10.957~MeV in $^{16}$O.  The 200-MeV data is from Ref.~[29].  The curves are
described in Fig.~1.
\endinsert

The choice of the momentum of the system depends on the most appropriate
model for the exchange in a (p,p$'$) transition.  Transitions where exchange is important
are the reactions to the $0^-$ states.  For the $0^+\to 0^-$ spin combination,
only the spin-longitudinal term in the interaction can contribute.  In a plane-wave
calculation, the analyzing power is non-vanishing only if the exchange is included in
finite range.  In order to handle this properly, the calculations were made with the
distorted-wave program DWBA86 [28].  Figure 2 shoes the same set of curves
as Fig.~1 for the
200-MeV cross section for the $T=0$, $0^-$ state at 10.957~MeV in $^{16}$O.
The data are from Ref.~[29].  The solid
($\vec P$ along $\theta /2$) curve is now substantially different from the medium dash
($\vec P$ along $\hat z$) curve.  (The normalization of the cross section is arbitrary
as electron scattering is not sensitive to this transition and there is no other
reference that precisely constrains the structure.)
This demonstrates that there is a sensitivity to the choice of
system momentum for
tansitions that depend on the treatment of exchange.

Similar tests were conducted for measurements of the polarization
transfer coefficients for unnatural parity transitions at 200~MeV [14].
In this case the formfactor for these transitions is peaked at a large
radius and Pauli blocking effects of any form are suppressed.
\vskip.3in
\noindent
{\bf V.~~Conclusions}
\medskip
As the basis for (p,p$'$) reaction calculations, we
have used a $G$-matrix that is
generated from an exact treatment of the angular dependence of the Pauli exclusion
operator.  For proton energies near 100~MeV, lifting the spherical averaging approximation
makes significant changes to the (p,p$'$) cross section and analyzing power.  The exact
treatment should be used whenever precise Pauli exclusion medium effects are considered.  This becomes less important as the projectile energy rises.

One consideration of exchange in the coordinate space distorted-wave calculation leads to
the choice of the system momentum that is not along the incident
projectile direction.
Transitions that are sensitive to tensor forces or finite-range exchange may be sensitive
also to the angle of this system momentum, even at the higher energies.
\bigskip
The authors acknowledge financial support from the U.S. National Science Foundation
under grant number PHY-0100348 (E.J.S.)
and the U.S. Department of Energy under grant number
DE-FG03-00ER41148 (F.S.).
\vskip.3in
\item{[1]} H.V. von Geramb, {\it The Interaction Between Medium Energy Nucleons in Nuclei
-- 1982,} AIP Conf.\ Proc.\ No.\ 97 (AIP, New York, 1983) p.\ 44.
\item{[2]} L. Rikus, K. Nakano, and H.V. von Geramb, Nucl.\ Phys.\ {\bf A414}, 413 (1984).
\item{[3]} K. Nakayama and W.G. Love, Phys.\ Rev.\ C {\bf 38}, 51 (1988).
\item{[4]} K. Amos, P.J. Dortmans, H.V. von Geramb, S. Karataglidis, and J. Raynal, Adv.\
Nucl.\ Phys.\ {\bf 25} 275 (2000).
\item{[5]} F. Sammarruca, E.J. Stephenson, and K. Jiang, Phys.\ Rev.\ C {\bf 60}, 064610
(1999).
\item{[6]} M.I. Haftel and F. Tabakin, Nucl.\ Phys.\ {\bf A158}, 1 (1970).
\item{[7]} T. Cheon and E.F. Redish, Phys.\ Rev.\ C {\bf 59}, 331 (1989).
\item{[8]} E. Schiller, H. M\"uther, and P. Czerski, Phys.\ Rev.\ C {\bf 59}, 2934 (1999);
{\bf 60}, 059901 (1999).
\item{[9]} K. Suzuki, R. Okamato, M. Kohno, and S. Nagata, Nucl.\ Phys.\ {\bf A665} 92
(2000).
\item{[10]} F. Sammarruca, X. Meng, and E.J. Stephenson, Phys.\ Rev.\ C {\bf 62}, 014614
(2000).
\item{[11]} A.A. Ioannides and R.C. Johnson, Phys.\ Rev.\ C {\bf 17}, 1331 (1978).
\item{[12]} E.J. Stephenson, C.C. Foster, P. Schwandt, and D.A. Goldberg, Nucl.\ Phys.\
{\bf A359}, 316 (1981).
\item{[13]} E.J. Stephenson, J.C. Collins, C.C. Foster, D.L. Friesel, W.W. Jacobs, W.P.
Jones, M.D. Kaitchuck, P. Schwandt, and W.W. Daehnick, Phys.\ Rev.\ C {\bf 28}, 134 (1983).
\item{[14]} F. Sammarruca, E.J. Stephenson, K. Jiang, J. Liu, C. Olmer, A.K. Opper, and
S.W. Wissink, Phys.\ Rev.\ C {\bf 61}, 014309 (1999).
\item{[15]} J.J. Kelly and S.J. Wallace, Phys.\ Rev.\ C {\bf 49}, 1315 (1994).
\item{[16]} F. Sammarruca and E.J. Stephenson, Phys.\ Rev.\ C {\bf 64}, 034006 (2001).
\item{[17]} R. Machleidt, Adv.\ Nucl.\ Phys.\ {\bf 19}, 189 (1989).
\item{[18]} K.A. Brueckner, C.A. Levinson, and H.M. Mahmoud, Phys.\ Rev.\ {\bf 95}, 217
(1954).
\item{[19]} H.A. Bethe, Phys.\ Rev.\ {\bf 103}, 1353 (1956).
\item{[20]} J. Goldstone, Proc.\ R. Soc.\ London, Ser.\ A {\bf 239}, 267 (1957).
\item{[21]} H.A. Bethe, Annu.\ Rev.\ Nucl.\ Sci.\ {\bf 21}, 93 (1971).
\item{[22]} V.G.J. Stoks {\it et al.},\ Phys.\ Rev.\ C {\bf 49}, 2950 (1994).
\item{[23]} W.G. Love and M.A. Franey, Phys.\ Rev.\ C {\bf 24}, 1073 (1981).
\item{[24]} H. Seifert {\it et al.},\ Phys.\ Rev.\ C {\bf 47}, 1615 (1993).
\item{[25]} H. Seifert, Ph.D.\ thesis, University of Maryland, 1990.
\item{[26]} James J. Kelly, Program Manual for LEA, 1995.
\item{[27]} T.N. Buti {\it et al.},\ Phys.\ Rev.\ C {\bf 33}, 755 (1986); and references
therein.
\item{[28]} R. Schaeffer and J. Raynal, program DWBA70; S. Austin, W.G. Love, J.R.
Comfort, and C. Olmer, extended version DWBA86 (unpublished).
\item{[29]} R. Sawafta, private communication.
\vfill
\end